\documentclass[12pt]{article}  
\usepackage{epsfig}

\def\ifmath#1{\relax\ifmmode #1\else $#1$\fi}%
\newcommand{\bit}{\begin{itemize}}
\newcommand{\eit}{\end{itemize}}
\newcommand {\ee}         {\mathrm{e}^+\mathrm{e}^-}
\newcommand {\ipb}   {\mbox{pb$^{-1}$}}
\newcommand{\xarc}{\ifmath{x_{\mathrm{arc}}}}
\newcommand{\Ebeam}{\ifmath{E_{\mathrm{beam}}}}
\newcommand{\ECM}{\ifmath{E_{\mathrm{CM}}}}
\newcommand{\Epol}{\ifmath{E_{\mathrm{pol}}}}
\newcommand{\Enmr}{\ifmath{E_{\mathrm{NMR}}}}
\newcommand{\Bnmr}{\ifmath{B_{\mathrm{NMR}}}}
\newcommand{\Bfl}{\ifmath{B_{\mathrm{FL}}}}

\newcommand  {\Qs}   {\ifmath{Q_{\mathrm{s}}}}
\catcode`@=11 
\def \gsim{\mathrel{\mathpalette\@versim>}}
\def \lsim{\mathrel{\mathpalette\@versim<}}
\def \@versim#1#2{\lower0.4ex\vbox{\baselineskip\z@skip\lineskip\z@skip
     \lineskiplimit\z@\ialign{$\m@th#1\hfil##\hfil$%
     \crcr#2\crcr\sim\crcr}}}
\catcode`@=12 
\newcommand{\beq}{\begin{equation}}
\newcommand{\eeq}{\end{equation}}

\begin{document}

\begin{titlepage}
{\center\large
EUROPEAN LABORATORY FOR PARTICLE PHYSICS \\
}
\vspace*{2mm}
\begin{flushright}
CERN-EP/98-191 \\
CERN-SL/98-073 \\
11 December, 1998 \\
Revised 19 March, 1999

\end{flushright}

\begin{center}
{\Large  \bf 
Evaluation of the LEP centre-of-mass energy \\
above the W-pair production threshold}

\vspace*{0.4cm}

The LEP Energy Working Group\\

\vspace*{0.6cm}

A.~Blondel$^{  1}$, 
M.~B{\oe}ge$^{  2a}$, 
E.~Bravin$^{  2}$, 
P.~Bright-Thomas$^{  2b}$,
T.~Camporesi$^{  2}$, 
B.~Dehning$^{  2}$, 
M.~Heemskerk$^{  2}$, 
M.~Hildreth$^{  2}$, 
M.~Koratzinos$^{  2}$, 
E.~Lan\c{c}on$^{  3}$,
G.~Mugnai$^{  2}$, 
A.~M\"uller$^{  2,4}$, 
E.~Peschardt$^{  2}$, 
M.~Placidi$^{  2}$, 
N.~Qi$^{  2c}$,
G.~Quast$^{  4}$, 
P.~Renton$^{  5}$, 
F.~Sonnemann$^{  2}$, 
E.~Torrence$^{  2}$, 
A.~Weber$^{  6}$,
P.~S.~Wells$^{  2}$,
J.~Wenninger$^{  2}$, 
G.~Wilkinson$^{  2}$

\end{center}

\begin{flushleft}
\footnotesize
$^{  1}$Laboratoire de Physique Nucl\'{e}aire et des Hautes Energies, 
Ecole Polytechnique,  IN$^2$P$^3$-CNRS,  F-91128 Palaiseau Cedex,  France\\
$^{  2}$CERN, European Organisation for Particle Physics,
CH-1211 Geneva 23, Switzerland \\
$^{  3}$CEA, DAPNIA/Service de Physique des Particules, CEA-Saclay, 
F-91191 Gif-sur-Yvette Cedex, France\\
$^{  4}$Institut f\"{u}r Physik,  Universit\"{a}t Mainz,  D-55099 Mainz, 
Germany \\
$^{  5}$Department of Physics,  University of Oxford,  Keble Road,  Oxford
OX1 3RH,  UK\\
$^{  6}$Aachen I (RWTH), I. Physikalisches Institut, 
Sommerfeldstrasse, Turm 28, D-52056 Aachen, Germany \\
$^{  a}$Now at: PSI - Paul Scherrer Institut, Villigen, Switzerland\\
$^{  b}$Now at: School of Physics and Astronomy, 
University of Birmingham, Birmingham B15 2TT, UK\\
$^{  c}$Now at: Institute of High Energy Physics,            
                    Academia Sinica,                
                    P.O. Box 918,    Beijing, China
\end{flushleft}

\vspace*{0.6cm}

\begin{abstract}

Knowledge of the centre-of-mass energy at LEP2 
is of primary importance to set the absolute energy scale
for the measurement of the W-boson mass.
The beam energy above 80~GeV is derived from continuous
measurements of the magnetic bending field by
16 NMR probes situated in a number of the LEP dipoles. 
The relationship between the fields measured
by the probes and the
beam energy is calibrated against 
precise measurements of the average 
beam energy between 41 and 55~GeV made using
the resonant depolarisation technique.
The linearity of the relationship is tested by
comparing the fields measured by the probes with the
total bending field measured by a flux loop.
This test results in the largest contribution to the systematic uncertainty.
Several further corrections are applied to derive
the the centre-of-mass energies
at each interaction point.
In addition the centre-of-mass energy spread is evaluated.
The beam energy 
has been determined
with a precision of 25~MeV for the data taken in 1997,
corresponding to a relative precision of $2.7\times 10^{-4}$.
This is small in comparison to the present uncertainty
on the W mass measurement at LEP. However, the ultimate
statistical precision on the W mass with the full LEP2 data sample
should be around 25~MeV, and a smaller uncertainty
on the beam energy is desirable. Prospects for 
improvements are outlined.
\end{abstract}

\vspace{4mm}     

\begin{center}
\large
Submitted to Eur.\ Phys.\ J.\ C.
\end{center}
\end{titlepage}

\section{Introduction}

The centre-of-mass energy of the large electron-positron (LEP) collider 
increased to 161~GeV in 1996, allowing W-pair production in $\ee$ 
annihilation for the first time. This marked the start of the LEP2
programme. The energy has been further increased in a series of
steps since then.
A primary goal of LEP2 is to measure the W-boson mass, 
$M_{\mathrm W} \approx 80.4$~GeV. 
The beam energy sets the absolute
energy scale for this measurement, leading to 
an uncertainty of 
$\Delta M_{\mathrm W} / M_{\mathrm W}  \approx
\Delta \Ebeam/ \Ebeam$.
With the full LEP2 data sample, the statistical uncertainty
on the W mass is expected to be around 25~MeV. To avoid a significant
contribution to the total error, this sets a target of
$\Delta \Ebeam/\Ebeam \approx 10^{-4}$,
i.e. 10 to 15~MeV uncertainty for a beam energy around 90~GeV.
This contrasts with 
LEP1, where the Z mass was measured with a total
relative uncertainty  of about $ 2 \times 10^{-5}$~\cite{ref:zpaper}.

The derivation of the centre-of-mass energy proceeds in 
several stages. First, the average
beam energy around the LEP ring is determined. 
The overall energy scale is normalised with respect to 
a precise reference in occasional dedicated measurements
during each year's running.
Time variations in the average beam energy are then taken into account.
Further corrections are 
applied to obtain the $\mathrm{e}^+$ and
$\mathrm{e}^-$ beam energies at the four interaction
points, and the centre-of-mass energy in the 
$\ee$ collisions. These procedures are elaborated below.

At LEP1, the average beam energy 
was measured directly at the physics operating energy
with a precision of better than 1~MeV by 
resonant depolarisation (RD)\cite{ref:zpaper}. 
The spin tune, $\nu$, determined by RD, is 
proportional to the beam energy averaged around the 
beam trajectory:
\beq
\nu = \frac{g_e-2}{2}\frac {\Ebeam}{m_e c^2}
\eeq
Both $\nu$ and \Ebeam\ are also 
proportional to the total integrated
vertical magnetic field, $B$, around the beam trajectory, $\ell$:
\beq
\Ebeam = \frac{e}{2\pi c} \oint_{\mathrm{LEP}}  B \cdot \mathrm{d} \ell
\eeq
Unfortunately, the RD technique can not be used in the LEP2
physics regime, because
depolarising effects increase sharply with beam energy, leading
to an insufficient build up of transverse polarisation to make
a measurement. In 1997, the highest energy measured by RD was 55~GeV.

The beam energy in LEP2 operation is therefore determined from an
estimate of the field integral derived from continuous
magnetic measurements by 16 NMR probes situated in some 
of the 3200 LEP main bend dipoles. These probes are read out during 
physics running and RD measurements, and they function at 
any energy above about 41~GeV. Although they 
only sample a small fraction
of the field integral, the relation between their readings
and the beam energy can be 
precisely calibrated against RD measurements in
the beam energy range 41 to 55~GeV.

The relation between the fields measured by the NMR probes and
the beam energy
is assumed to be linear, and to be valid up to
physics energies. Although the linearity can only be tested over
a limited range with the RD data themselves,
a second comparison of the NMR readings with
the field integral is available. A flux loop is installed in
each LEP dipole magnet, and provides
a measurement of 96.5\% of the field integral.
Flux loop experiments 
are performed only occasionally, without beam in LEP.
The change in flux is measured during a dedicated cycling
of the magnets. 
The local dipole bending fields measured by the
NMR probes are read out at several
steps in the flux-loop cycle, over the full range from
RD to physics energies. This provides an independent test of
the linearity of the relation between the probe fields
and the total bending field. 

The use of the NMR probes to transport the precise energy scale
determined by RD to the physics operating energy is the 
main novelty of this analysis. The systematic errors on the
NMR calibration are evaluated from the reproducibility of
different experiments, and the variations from probe to probe.
The dominant uncertainty comes from the 
quality of the linearity test with the flux loop.

At LEP2, with 16 NMR probes in the LEP tunnel, 
time variations in the 
dipole fields provoked by leakage 
currents from neighbourhood electric trains 
and due to temperature effects can be accounted for directly.
This is in contrast to the 
LEP1 energy measurement, where understanding the time 
evolution of the dipole fields during a LEP fill 
formed a major part of the analysis~\cite{ref:zpaper}.

The NMR probes and the flux loop measure only the 
magnetic field from the LEP dipoles, which is 
the main contribution to the field integral.
The LEP quadrupoles also
contribute to the field integral when the beam passes through them
off axis, which occurs if for any reason the beam is 
not on the {\it central orbit}.
The total orbit length is fixed by the RF accelerating frequency.
Ground movements, for example due to earth tides or 
longer time scale geological effects, move the LEP magnets
with respect to this fixed orbit~\cite{ref:zpaper}. 
At LEP2, deliberate changes in the RF frequency away from the
nominal central value 
are routinely used to optimise the luminosity by 
reducing the horizontal beam size.
This can cause occasional abrupt changes in the beam energy.
Orbit corrector magnets also make a small contribution to the
total bending field. 
All of these corrections
must be taken into account both when comparing the NMR measurements
with the RD beam energies, and in deriving the centre-of-mass
energy of collisions as a function of time.

The exact beam energy at a particular location differs from the average
around the ring 
because of the loss of energy by synchrotron radiation in the
arcs, and the gain of energy in the RF accelerating sections;
the total energy lost in one revolution is about 2~GeV at LEP2.
The $\mathrm{e}^+$ and
$\mathrm{e}^-$ beam energies at each interaction point are calculated 
taking into account the exact accelerating RF configuration. 
The centre-of-mass energy at the collision point can also be
different from the sum of the beam energies due to 
the interplay of collision offsets and dispersion. 
The centre-of-mass energies for each interaction point are calculated
every 15 minutes, or more often if necessary, 
and these values are distributed to the LEP experiments.

In the following section, the data samples and magnetic
measurements are described. The beam energy model is outlined in 
section~\ref{sec:model}. The calibration of the NMR probes 
and the flux-loop test are described in section~\ref{sec:nmrcal}.
More information on corrections to the beam energy from
non-dipole effects is given in
section~\ref{sec:quad}, and on IP specific corrections
to derive the centre-of-mass energy 
in section~\ref{sec:ipcor}. The systematic uncertainties for the
whole analysis are summarised in section~\ref{sec:syst}.
The evaluation of the instantaneous spread in centre-of-mass
energies is given in section~\ref{sec:espread}. In the conclusion,
the prospects for future improvement are also outlined.

\section{Data samples}
\label{sec:data}

\subsection{Luminosity delivered by LEP2}

LEP has delivered about 10~\ipb\ at each of two centre-of-mass
energies, 161 and 172~GeV, in 1996, and over 50~\ipb\ at a 
centre-of-mass energy of around 183~GeV in 1997. Combining
the data from all four LEP experiments, these data give
a measurement of the W mass with a precision of 
about 90~MeV~\cite{ref:thomson}. This paper emphasises
the 1997 energy analysis, with some information for 1996 
where relevant.

\subsection{Polarisation measurements}

\begin{table}[htb]
\begin{center}
\begin{tabular}{l|c|c|c|c|c|c}
\hline
Date     & Fill   & 41 GeV & 44 GeV & 50 GeV & 55 GeV  & Optics \\
\hline
19/08/96 & 3599  &        &     & yes &    & 90/60 \\
31/10/96 & 3702  &        & yes &     &    & 90/60 \\
03/11/96 & 3719  &        & yes & yes &    & 90/60 \\
\hline
17/08/97 & 4000  &        & yes &             &    & 90/60\\
06/09/97 & 4121  &        & yes & yes &         & 60/60\\
30/09/97 & 4237  &        & yes & yes &         & 60/60\\
02/10/97 & 4242  & yes & yes & yes & yes   & 60/60\\
10/10/97 & 4274  &        & yes &          &       & 90/60\\
11/10/97 & 4279  & yes & yes & yes & yes   & 60/60\\
29/10/97 & 4372  & yes & yes &        &         & 60/60 \\
\hline
\end{tabular}
\end{center}
\caption{Fills with successful polarisation measurements in 1996
and 1997. Each calibrated energy point is marked ``yes''.}
\label{tab:pol}
\end{table}

The successful RD experiments in 1996 and 1997 
are listed in table~\ref{tab:pol}.
To reduce uncertainties from fill-to-fill variations,
an effort was made to measure as many 
beam energies as possible with RD during
the same LEP fill.
Measuring two energies in the same fill was first
achieved at the end of 1996.
The need for more RD measurements 
motivated the ``k-modulation'' programme to measure the
offsets between beam pick-ups and quadrupole 
centres~\cite{ref:kmod},
the improved use of magnet position surveys,
and the development
of a dedicated polarisation optics~\cite{ref:jan}
(the 60/60 
optics\footnote{The optics are designated by the 
betatron advance between focusing quadrupoles of the
LEP arcs in the horizontal/vertical planes respectively.}).
These were all used in 1997~\cite{ref:bernd}. 
Improving the orbit quality and reducing depolarising
effects in this way resulted in
5 fills with more than one energy point, and
2 fills with 4 energy points, which allow a check
of the assumption that the measured
magnetic field is linearly related to the average
beam energy.
The range over which tests can be made
increased from 5 to 14~GeV between 1996 and 1997, 
and the maximum RD calibrated
energy increased from 50 to 55~GeV. 
At least a 4--5\% level of polarisation is needed
to make a reliable measurement, but only 
2\% level of polarisation was observed at 60~GeV in 1997.

\subsection{Magnetic measurements}

\begin{figure}[htb]
\centering
\mbox{\epsfxsize=0.8\textwidth
\epsfbox{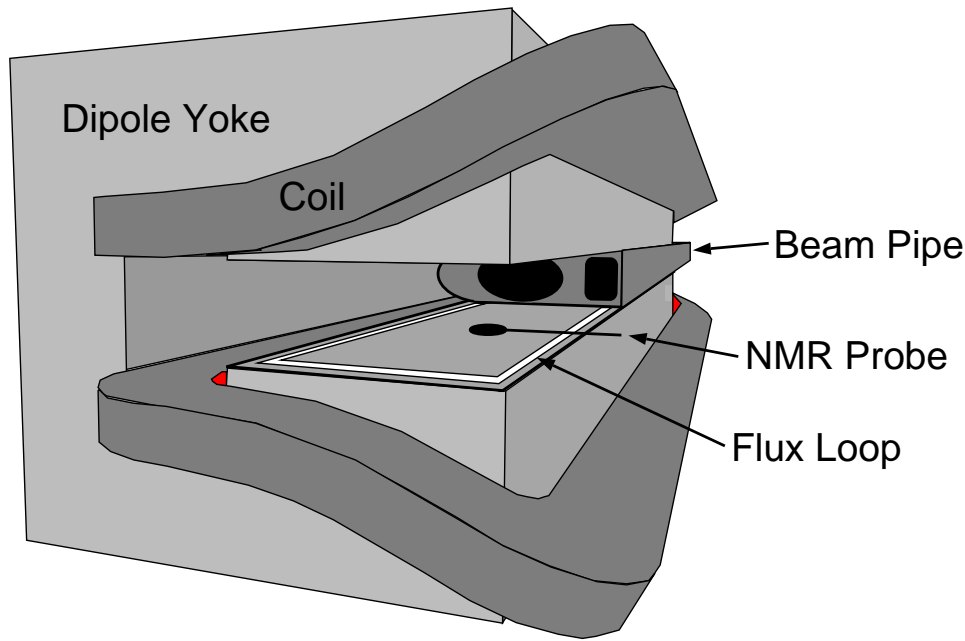} }
\caption{A LEP dipole magnet showing the flux
loop and an NMR probe.}
\label{fig:nmrfl}
\end{figure}

The LEP dipole fields are monitored continuously by NMR probes,
and in occasional dedicated measurements by the flux loop.
A total of 16 probes was installed for the 1996 LEP
run. The probes are positioned inside selected main bend dipoles,
as indicated in figure~\ref{fig:nmrfl}. Each octant has at least one
probe, and octants 1 and 5 have strings of
probes in several adjacent dipoles (and in one 
instance two probes in the same dipole). The probes measure the local 
magnetic field
with a precision of around $10^{-6}$, and they can
be read out about every 30 seconds.
Each probe only
samples the field in a small region of one out of 3200 dipoles.
A steel field plate is installed between each probe and
the dipole yoke to improve
the uniformity of the local magnetic field.
During normal physics running and RD measurements, 
the probe readings over five minute
time intervals are averaged. This reduces the
effect of fluctuations in the magnetic fields induced by
parasitic currents on the beam pipe (see section~\ref{sec:model}).
The probes are also read out during flux-loop measurements, as
described below.

In 1995 there were only two probes
in dipoles in the LEP ring, and prior to 1995, 
only a reference magnet powered
in series with the main dipoles was monitored~\cite{ref:zpaper}. The larger
number of available probes in 1996 and 1997 
has allowed a simplification of the 
treatment of the dipole magnetic field evolution during a fill
(see section~\ref{sec:model}).

The performance of the probes is degraded by synchrotron 
radiation at LEP2. A new probe gives a strong enough signal
to lock on to and measure for fields corresponding 
to beam energies above about 40~GeV.
During a year's running, this minimum measurable field gradually
increases, and the probes eventually have to be replaced. However,
if a stable frequency lock is achieved, then the value of the field measured
is reliable; only the range of measurable fields is compromised.
During 1996 and 1997, all of the probes were working at high energy,
with the exception of two probes, one in each multi-probe
octant, which were not available for the running at 172~GeV centre-of-mass
energy. 

The flux loop is also shown schematically in figure~\ref{fig:nmrfl}. 
In contrast to the NMR probes, the flux loop samples
98\% of the field of each main bend dipole,
excluding fringe fields at the ends, corresponding to
96.5\% of the total bending field of LEP.
The loop does not include the weak 
(10\% of normal strength) dipoles
at the ends of the arcs, the double-strength 
dipoles in the injection regions, or 
other magnets such as quadrupoles or
horizontal orbit correctors.

The flux loop measures the change in flux during
a dedicated magnet cycle, outside physics running,
and the corresponding change in the average main dipole
bending field, \Bfl, is calculated.
Five of these measurements were made in 1997. 
The gradual increase of the magnetic field during the
cycle is stopped for several minutes 
at a number of intermediate field values,
each of which corresponds to a known nominal beam energy, to
allow time for the NMR probes to lock and be read out.
The values chosen include the 
energies of RD measurements and physics running.
The average of the good readings for each probe is calculated
for each step.
These special
flux-loop measurements are referred to by an adjacent LEP
fill number (one measurement in fills 4000, 4121 and 4206 and
two measurements in fill 4434).
This possibility to cross-calibrate the field measured
by the flux loop and by the NMR probes is crucial to the analysis.

\section{The beam energy model}
\label{sec:model}

The LEP beam energy is calculated as a function of time 
according to the following formula:
\begin{eqnarray}
\Ebeam (t) & = &  (E_{\mathrm{initial}} + \Delta E_{\mathrm{dipole}}(t) )
 \label{eq:model}\\
& &
 \cdot  (1+C_{\mathrm{tide}} (t))\cdot  (1+C_{\mathrm{orbit}}) 
\cdot (1 + C_{\mathrm{RF}}(t)) \nonumber \\
& &
 \cdot  (1+C_{\mathrm{h.corr.}} (t))\cdot  (1+C_{\mathrm{QFQD}} (t)) \nonumber  . 
\end{eqnarray}

The first term, $E_{\mathrm{initial}}$, is the energy 
corresponding to the dipole field integral at the point 
when the dipoles reach operating conditions, i.e. 
after the beams have been ``ramped'' up to physics energy,
and after any bend modulation has been performed (see below).
For RD fills, $E_{\mathrm{initial}}$ is
calculated after each ramp to a new energy point.
The shift in energy caused by changes in the bending dipole
fields during a fill is given by $\Delta E_{\mathrm{dipole}}(t)$.

Both of these ``dipole'' terms are averages over the energies 
predicted by each functioning NMR probe. 
For the initial energy, equal weight
is given to each probe, since each gives an independent estimate
of how a magnet behaves as the machine is ramped to physics energy.
However, for the change in energy during a fill, equal weight is 
given to each octant. This gives a more correct average over
the whole ring for dipole rise effects provoked
by temperature changes, and by parasitic electrical currents on
the beam pipe caused by trains travelling in the neighbourhood.

Modelling the dipole energy using the 16 NMR probes has 
simplified the treatment compared to LEP1,
where only two probes
were available in the tunnel in 1995, and none in earlier years.
For LEP1, $E_{\mathrm{initial}}$ was derived from comparisons
with RD measurements, and the rise in a fill from a model of the
train and temperature effects (see~\cite{ref:zpaper}). 

The dipole rise effects are minimised by bend modulation,
i.e. a deliberate small amplitude variation of the dipole 
excitation currents after the end of the ramp\cite{ref:zpaper}. 
These were not recommissioned for the 1996 running, but in 1997
were carried out routinely for physics fills from fill 3948 on 5 August.

The remaining terms correct for other contributions to the
integral bending field, and are listed below.
They are discussed
in section~\ref{sec:quad}, and in more detail 
in reference~\cite{ref:zpaper}.
These terms must also all be taken into account 
when comparing the energy measured
at a particular time 
by RD with the magnetic field measured in the
main dipoles by the NMR probes. 
\begin{description}
\item[$C_{\mathrm{tide}} (t)$:] 
This accounts for the 
effect of earth tides which change the size of the LEP ring, 
effectively moving the quadrupole magnets with respect
to the fixed-length
beam orbit.
\item [$C_{\mathrm{orbit}}$:] This is evaluated once for each LEP fill.
It corrects for distortions of the ring geometry on a longer time scale
according to the measured average horizontal orbit displacement.
\item [$C_{\mathrm{RF}}(t)$:] 
Regular changes in the RF frequency away from the nominal central
frequency are made to optimise the luminosity, leading to this correction.
\item [$C_{\mathrm{h.corr.}} (t)$:] This accounts for
changes in the field integral from horizontal orbit corrector magnets
used to steer the beam.
\item [$C_{\mathrm{QFQD}} (t)$:] Stray fields are 
caused when different excitation currents are
supplied to focussing and defocussing quadrupoles in the LEP lattice.
These are taken into account by this term.
\end{description}

\section{Calibration of NMR probes}
\label{sec:nmrcal}

\subsection{Calibration of NMR probes with RD measurements}
\label{sec:nmrpol}

The magnetic fields $\Bnmr^i$ measured by each NMR $i=1,16$,
are converted into an equivalent beam energy. The 
relation is assumed to be linear, of the 
form
\beq
\label{eq:epol}
 \Enmr^i = a^i + b^i \Bnmr^i.
\eeq
In general, the beam energy is expected to be
proportional to the integral bending field. 
The two parameters for each probe are
determined by a combined fit to all of the energies measured
by resonant depolarisation. 
The NMR probes only give
an estimate of the dipole contribution to the integral
bending field, so all the other effects, such as those
due to coherent quadrupole motion, must be taken into account
according to equation~\ref{eq:model} in order
to compare with the energy measured by RD. 
A further complication arises because two different weighted
averages over probes are used to derive $E_{\mathrm{initial}}$
and  $\Delta E_{\mathrm{dipole}}(t)$, so in practice an iterative
procedure is used. 
The average offset, $a$, is 27~MeV, with an rms spread
over 16 NMR probes of 64~MeV. The average slope, $b$, is
91.17~MeV/Gauss, with an rms spread of 0.25~MeV/Gauss over 16 probes.

\begin{figure}[htb]
\centering
\mbox{
\mbox{\epsfxsize=0.5\textwidth
\epsfbox{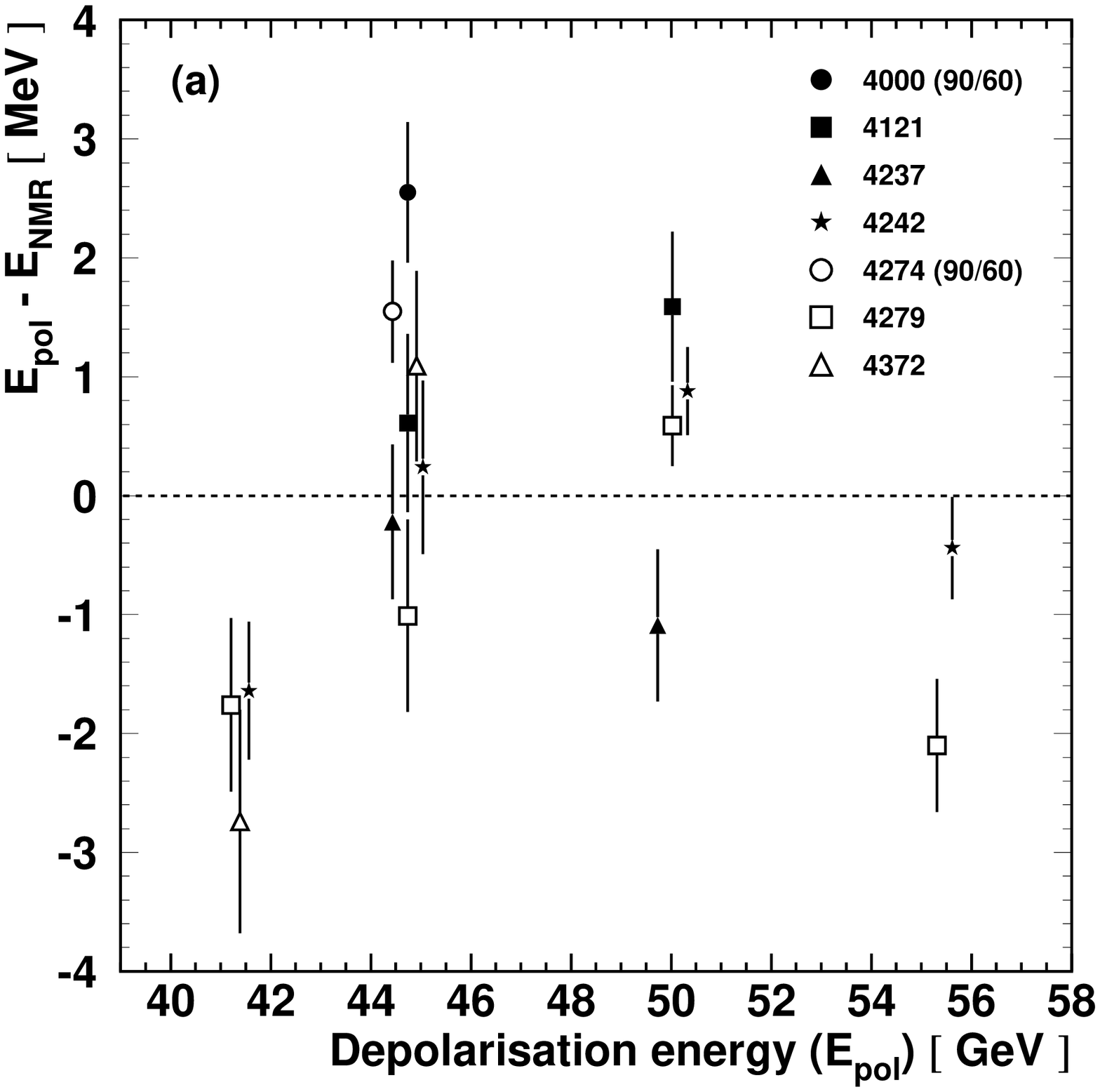} }
\mbox{\epsfxsize=0.5\textwidth
\epsfbox{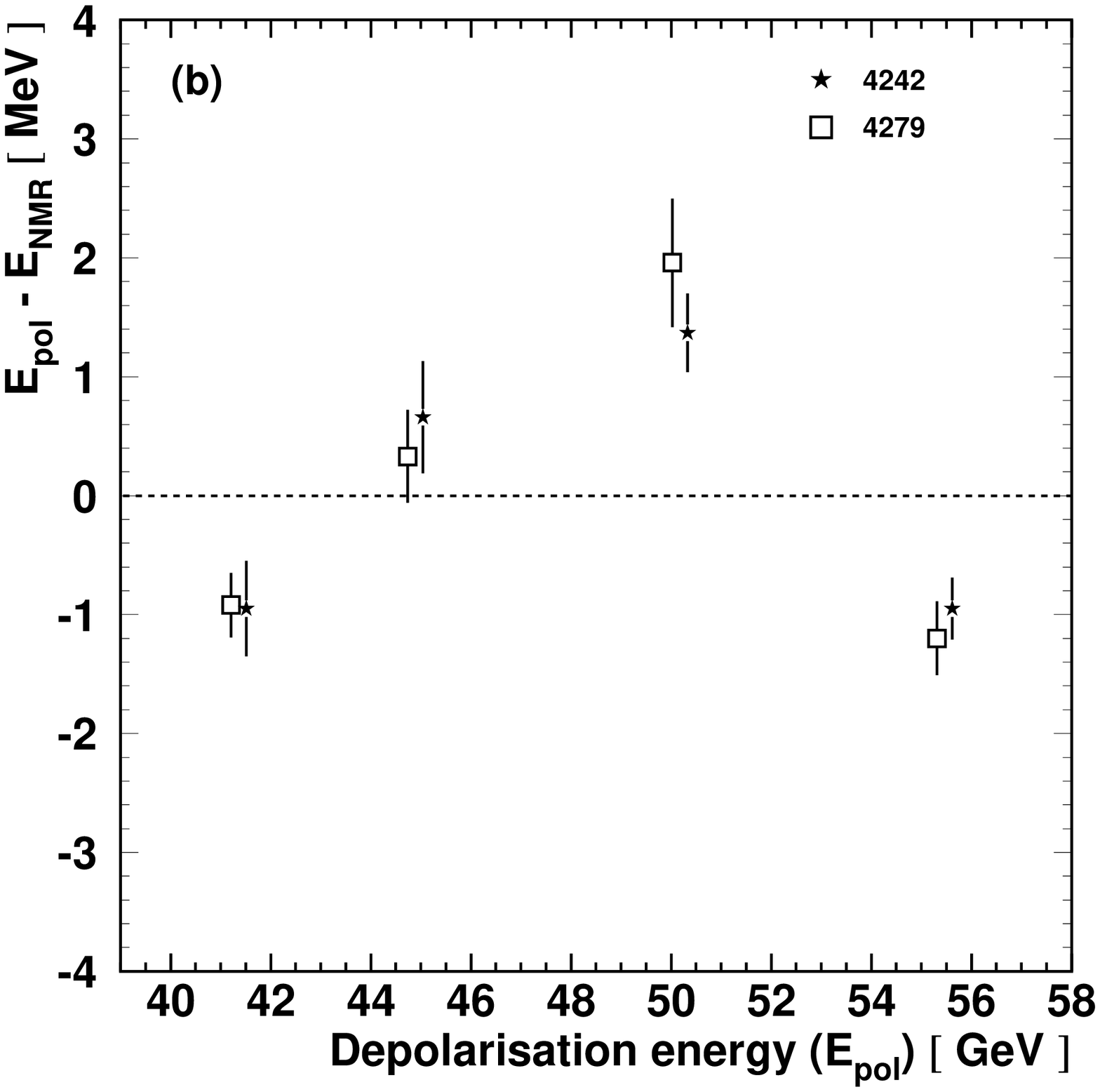} }
}
\caption{Residuals of the fit comparing RD energies to the energies
predicted by the model (a) for a simultaneous fit to all fills
and (b) for individual fits to each four point fill. 
For clarity, the 
points for different fills have been plotted at slightly 
displaced depolarisation energies.}
\label{fig:allresid}
\end{figure}

The residuals, $\Epol - \Enmr^i$, are examined for each NMR.
The residuals evolve with beam energy in a 
different way for different probes, 
but for a particular probe
this behaviour is reproduced from fill to fill.
The residuals averaged over NMR probes at each polarisation point 
are shown in figure~\ref{fig:allresid}(a),
in which the errors are displayed as the rms/$\sqrt{N}$, where $N\le 16$
is the number of NMR probes functioning for the measurement.
This figure shows the average residuals with respect to the simultaneous fit
to all polarisation fills in 1997, which was used to calibrate the
NMR probes. In figure~\ref{fig:allresid}(b), the residuals for the 
two fills with four RD energies are shown. Here the fit is made
to each fill individually. The residuals show a reproducible small
but statistically significant deviation from zero, with the 
45 and 50~GeV points being a few MeV higher than those at
41 and 55~GeV. Despite some fill-to-fill scatter, this shape is
present in all fills, not just the two fills with four RD energies.

\subsection{Predicted energy for physics running}

\label{sec:ephys}

Using the calibration coefficients determined in section~\ref{sec:nmrpol},
the magnetic field measured by each NMR during physics running can be
used to predict the beam energy.
To assess the variations over the NMR probes, 
the average magnetic field is calculated for each NMR over
all of the physics running at a
nominal beam energy of 91.5~GeV. The average physics energies
derived from these average fields have an
rms scatter over the NMR probes of about 40~MeV,
contributing $40/\sqrt{16}=10$~MeV
to the systematic uncertainty from the normalisation procedure.

In 1996, the limited number of available 
RD measurements were fitted to a line
passing through the origin, $\Epol = p^i \Bnmr^i$. If this is
tried for the 1997 data, the rms scatter increases to 60~MeV 
and the central value shifts by around 20~MeV. This is taken
into account in evaluating the uncertainty for the 1996 data,
as discussed in section~\ref{sec:e96err}.
The reduced scatter for the two parameter fit can be taken to
imply that a non-zero offset improves the description of the
energy-magnetic field relation, or that it is an advantage
to impose the linearity assumption only over the region
between polarisation and physics energies.

Using different polarisation 
fills as input to the procedure gives some variation. 
Fill 4372 has the most unusual behaviour. This is partly because it
only samples the two lower energy points, which have a different
average slope to the four point fills. This fill also has the 
smallest number of functioning NMR probes, since it is at the end of the
year. It therefore has little weight in the overall average. 
An uncertainty of 5~MeV is assigned to cover the range of 
central values derived using different combinations of polarisation
fills.

The average residuals of the fit show a characteristic shape 
which is a measure of the non-linearity in the beam energy 
range 41--55~GeV. 
The amplitude of the deviations is larger than the statistical scatter
over the NMR probes.
The linearity is best examined by making fits to the individual 4 point fills.
If the errors are inflated to achieve a 
$\chi^2$/dof of~1, then they imply an uncertainty at physics energy of 7~MeV
for the linear extrapolation.
No additional uncertainty is included to account for this observation,
because it is covered by the larger uncertainty assessed in 
section~\ref{sec:fltest}, where the linearity assumption is tested by
a comparison of NMR and flux-loop measurements.

\subsection{Initial fields for physics fills}
\begin{figure}[b]
\centering
\mbox{\epsfxsize=0.6\textwidth
\epsfbox{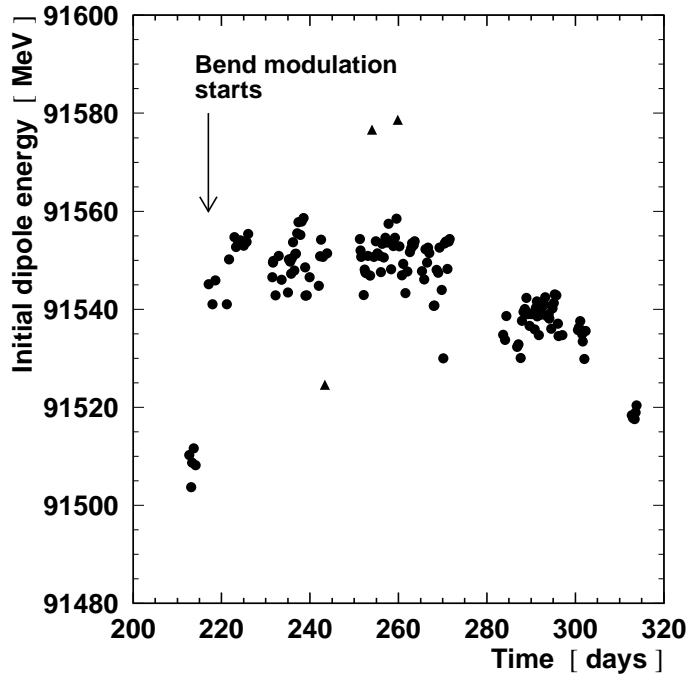} }
\caption{ Initial dipole field for all 183~GeV fills. The fills 
marked with a triangle have anomalous initial currents in the
main bend dipoles. A bend modulation was carried out at
the start of each fill from 5 August (day 217).}
\label{fig:inidip}
\end{figure}

The estimates of $E_{\mathrm{initial}}$, the initial energy
from the dipole contribution to the bending field, 
for all 
fills with a nominal centre-of-mass energy of 183~GeV are shown
in figure~\ref{fig:inidip}. 
There is a change in initial field after 5 August, attributed 
to the implementation of bend modulation at the start of each fill.
A small 
drift in the dipole excitation current for the same nominal setting
was observed during the year,
and one or two fills have anomalous initial values due to 
known incorrect setting of the excitation current.
The overall rms spread in $E_{\mathrm{initial}}$ is 11~MeV for 148 fills.
The error on the mean beam energy from these variations and from 
other rare 
anomalies at the starts of fills is taken to be 2~MeV.

\subsection{Uncertainty due to dipole rise}

The average of the NMR probes is used to take into account changes
in energy due to temperature and parasitic currents which
cause the dipole bend field to change.
Bend modulation~\cite{ref:zpaper} was not
performed in 1996, but in 1997 
was routinely carried out at the start of each
fill from 5 August (fill 3948). The average rise in
dipole field during a fill was therefore smaller 
in 1997 than in 1996:
the total dipole rise effect on the average beam energy was only 3.5~MeV.
From experience at LEP1, and the fact that the 16 NMR probes give a good
sampling of the whole ring, the uncertainty is expected to be
less than 25\% of
the effect, so 1~MeV is assigned as the uncertainty due to the dipole rise.

\subsection{Test of NMR calibration using the flux loop}
\label{sec:fltest}

\begin{figure}[b]
\centering
\mbox{
\mbox{\epsfxsize=0.5\textwidth
\epsfbox{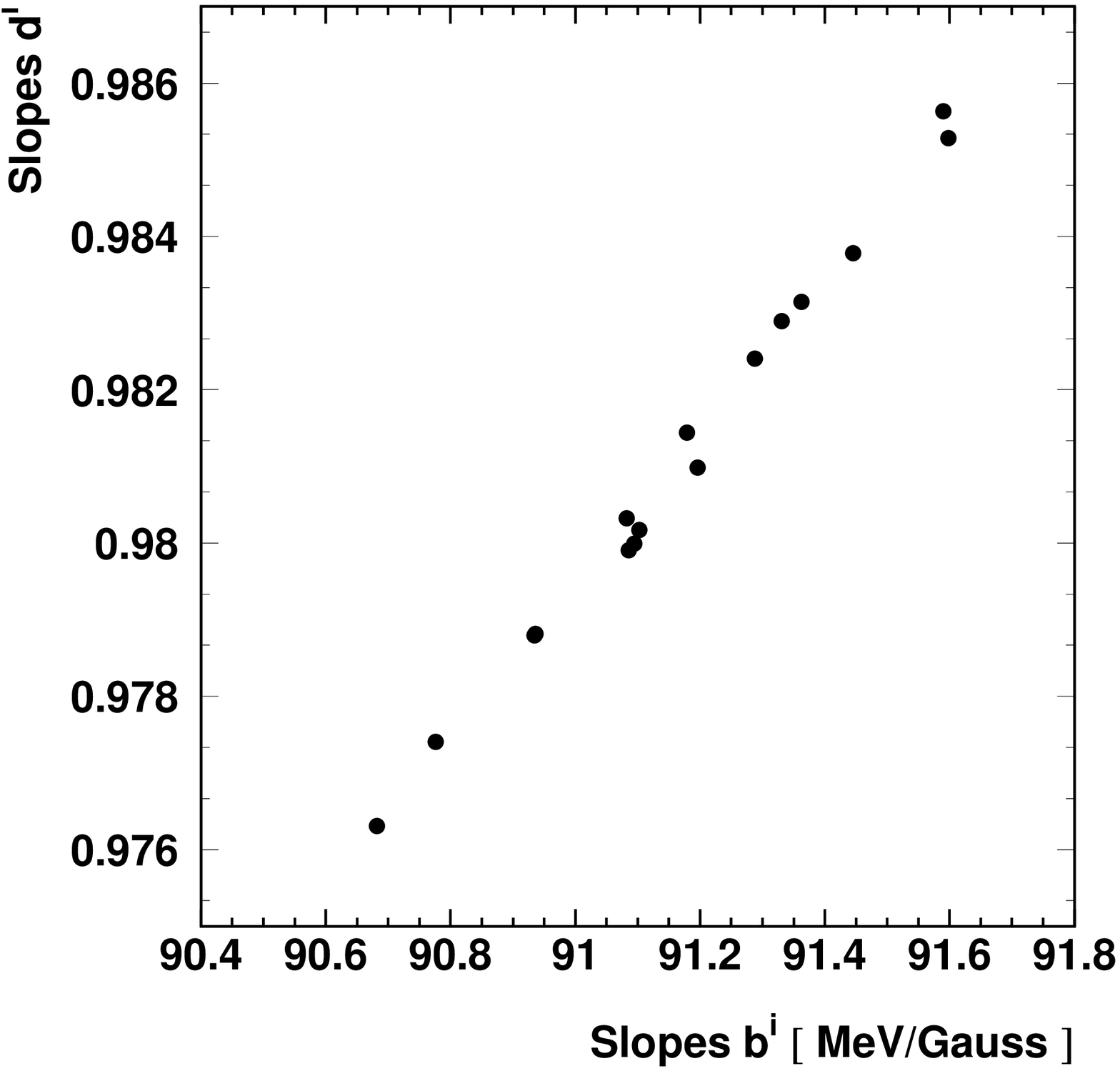} }
\mbox{\epsfxsize=0.5\textwidth
\epsfbox{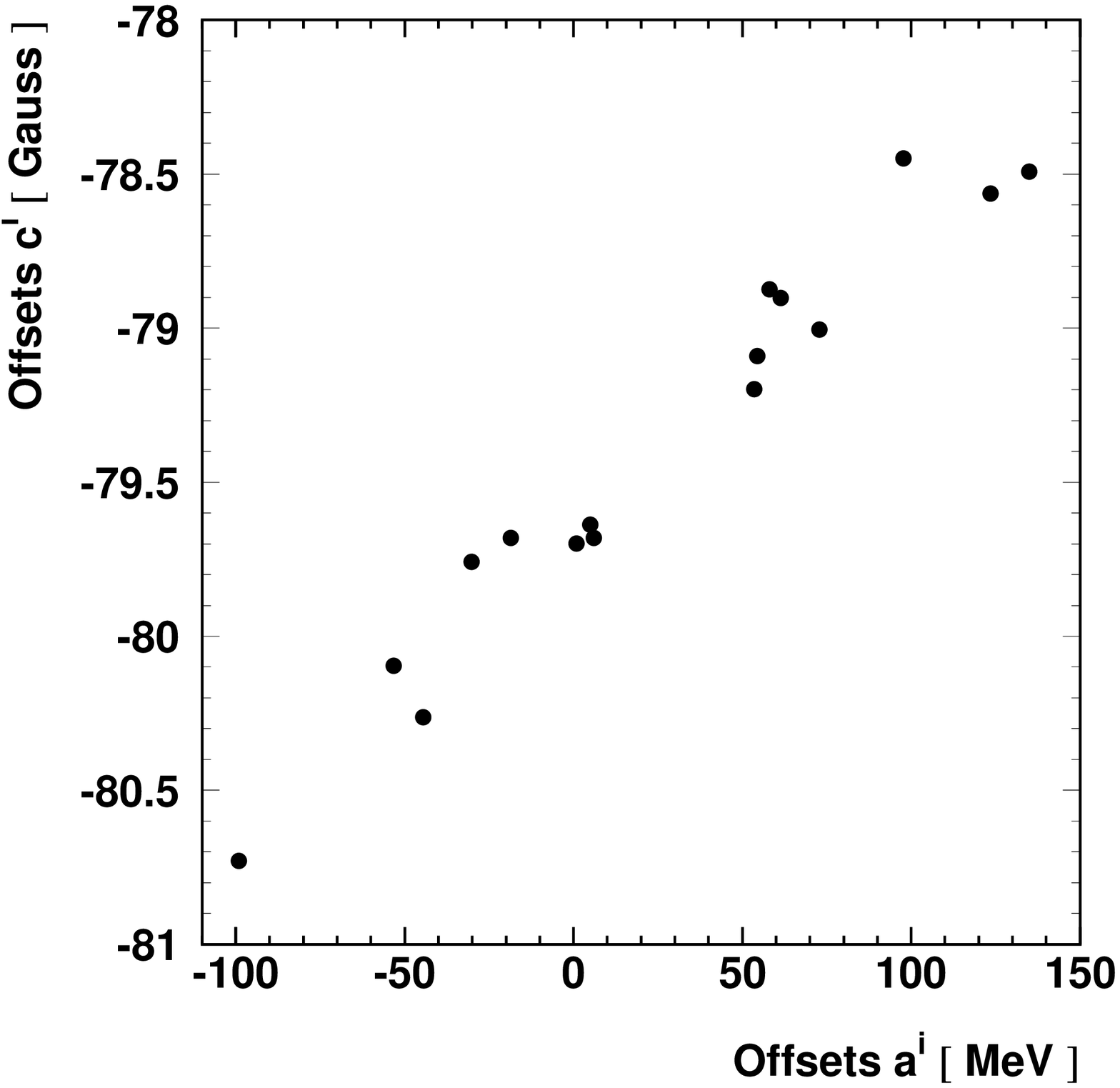} }
}
\caption{The slopes and offsets of 
equations~\ref{eq:correlab} and \ref{eq:correlcd} 
comparing the field measured by each NMR probe
with the RD and flux-loop measurements.
The values shown are averages over the 
five flux-loop measurements. There is one entry per NMR probe 
in each plot.}
\label{fig:correl}
\end{figure}

The estimate of the integral dipole field from the NMR probes can also
be compared with the measurement of 96.5\% of the total bending
field by the flux loop. This allows a test of the entire 
extrapolation method.
From a fit in the 41--55~GeV region, 
the NMR probes can be used to predict the average bending field
measured by the flux loop at the setting corresponding to physics
energy.
If the NMR probes can predict the flux-loop field, and the beam
energy is proportional to the total bending field, then it is a 
good assumption that the probes are also able to predict the 
beam energy in physics. The flux loop can not be used to predict
the beam energy in physics directly, since neither the slope nor the offset
of the relationship between measured field and beam energy are known
with sufficient precision. However, 
each point in the flux loop does correspond
to a specific setting of the
nominal beam energy. The flux-loop measurement is from 7 to 100~GeV.

For the test to be valid, a strong correlation should be observed
between the offsets $a^i$ and $c^i$, 
and between the slopes $b^i$ and $d^i$,
from the fits of the field measured by each probe, $i$,
to the polarisation and flux-loop data:
\begin{eqnarray}
\label{eq:correlab}
 \Epol  =  a^i + b^i \Bnmr^i &&\mbox{and}\\
\label{eq:correlcd}
 \Bfl   =  c^i + d^i \Bnmr^i && \mbox{fit restricted to 41--55 GeV}.
\end{eqnarray}
The fitted parameters for each NMR 
are shown in figure~\ref{fig:correl}, and the expected
correlation is seen.
The average offset, $c$, is $-79.38$~Gauss, 
with an rms spread over the 16 values of 0.67. This
offset corresponds to the 7~GeV 
nominal beam energy setting
at the start of the flux-loop cycle.
The average slope, $d$, is 0.9811, with an rms spread over
16 NMR probes of 0.0027. The field plates cause the
slope to be 2\% different from unity.

The residuals with respect to equation~\ref{eq:correlcd} above the
fit region (41--55~GeV) are used to test the linearity assumption.
The residual difference 
in Gauss between the flux-loop and NMR fields at a particular beam energy
can be converted to a residual bias in MeV by a scale factor of
92.9 MeV/Gauss (corresponding to the ratio of average slopes, $b/d$).

\begin{figure}[p]
\centering
\mbox{\epsfxsize=0.565\textwidth
\epsfbox{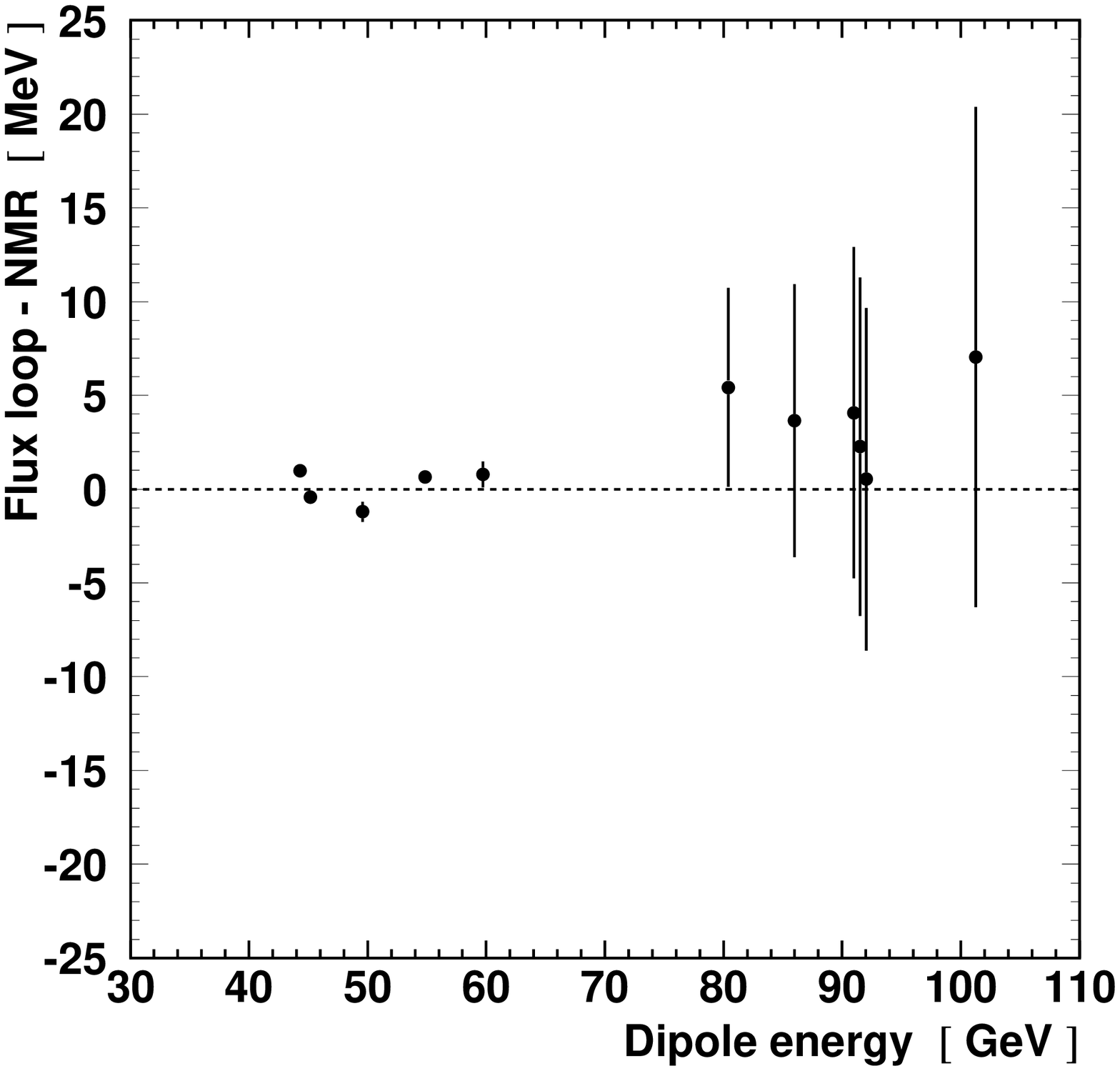} }
\caption{The difference scaled by 92.9 MeV/Gauss
between the magnetic field measured by the 
flux loop and predicted by the NMR probes
as a function of the nominal beam energy for fill 4000.
The error bars give the rms scatter over the probes, 
divided by the square root of the number of probes.}
\label{fig:fluxa}
\begin{center}
\mbox{\epsfxsize=0.565\textwidth
\epsfbox{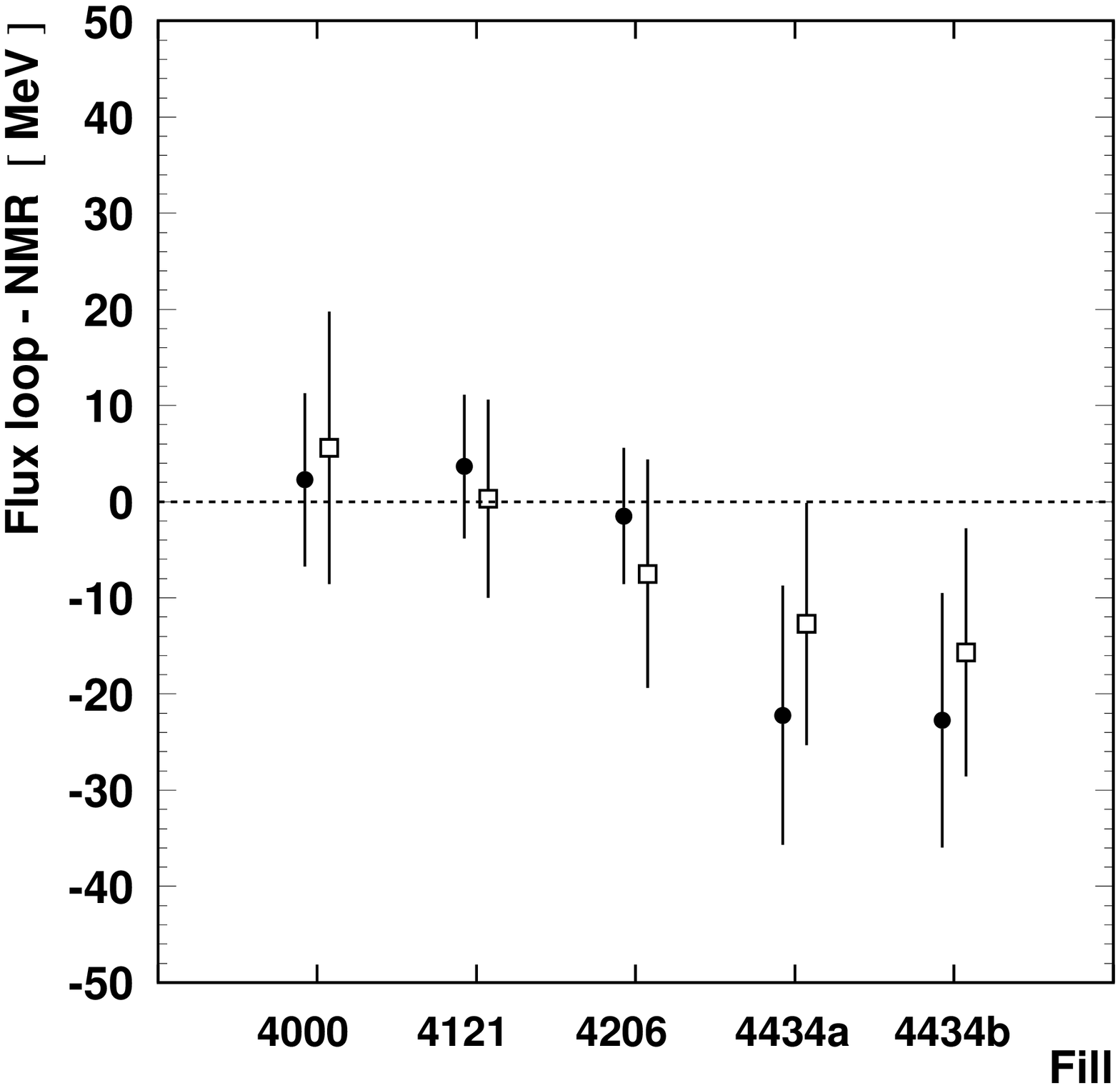} }
\end{center}
\caption{The difference scaled by 92.9 MeV/Gauss 
between the magnetic field
measured by the flux loop and predicted by the NMR probes
at physics energy (91.5~GeV)
for each flux-loop measurement. The solid points are for all
working probes for each fill, while the empty points are
for the common set of probes working for all fills.}
\label{fig:fluxb}
\end{figure}

An example fit to one flux-loop measurement is shown 
in figure~\ref{fig:fluxa}. The average over the NMR probes
of the scaled residuals to the fits
to equation~\ref{eq:correlcd} are shown at each nominal
beam energy.
The error bars
show the rms scatter divided by the square root of the number
of working probes.
The fit is in the range 41--55~GeV, and the error bars
increase above this region.
The deviations measured at the 
physics energy of 91.5~GeV for each of the five 
flux-loop measurements are shown in figure~\ref{fig:fluxb}, 
using the same conventions. 
The probes in different magnets show a different evolution of
the residuals as the nominal beam energy increases. However, 
for a particular magnet the behaviour is similar for each
flux-loop measurement.

The average bias
at physics energy is up to $20$~MeV, with an rms
over the probes of 30--40~MeV, corresponding to an uncertainty
of 10--20~MeV depending on the number of working probes.
The size of the bias tends to increase during the year; the
bias becomes more negative. 
This is partially understood 
as being due to the smaller sample of NMR probes available in the latter
part of the year, as is also illustrated in figure~\ref{fig:fluxb}.
To account for the correlated uncertainties from measurement
to measurement, the difference in bias between the first and
last measurements has been found for all of the probes that
are common to the two. A significant average difference
of $-21\pm5$~MeV is observed, for which no explanation has
been found. In fact, only the last two
flux loops include a 41~GeV point, but the biases measured are
not sensitive to excluding this point altogether.
Other systematic effects that have been observed, for example
a discrepancy of around 2~MeV if two very close by energy points are
measured, are too small to explain the trend.

\begin{figure}[bht]
\begin{center}
\mbox{\epsfxsize=0.6\textwidth
\epsfbox{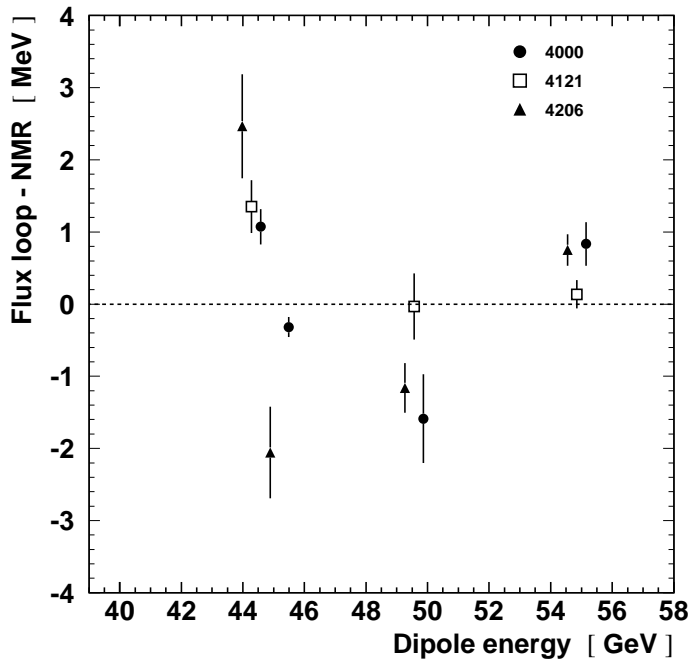} }
\end{center}
\caption{The residuals in the low energy 
   region with respect to separate fits to each of the 
flux-loop measurements 4000,4121 and 4206, using the subset
   of NMR probes available at all points shown in all
   3 fills. Too few NMR probes are working for all low energy points to 
include fill 4434 here.
}
\label{fig:fluxc}
\end{figure}

A detailed comparison of NMR and flux-loop data in the region of the RD
measurements (41--55~GeV), shown in figure~\ref{fig:fluxc},
reveals a different non-linearity to the NMR--polarisation
comparison. However, the subset of probes included here is not
exactly the same as in figure~\ref{fig:allresid},
and the average residuals to the flux-loop 
fits are only a few MeV
which is at the limit of the expected precision.

The tests with the flux loop are not used to correct the NMR 
calibration from polarisation data, but are taken as 
an independent estimate
of the precision of the method. A systematic uncertainty of 20~MeV
covers the maximum difference
seen during the year. An average over the flux-loop measurements would
give a smaller estimate.

\subsection{Uncertainty from bending field outside the flux loop}

The flux loop is embedded in each of the main LEP bending dipoles, 
but only samples 98\% of the total bending field of each dipole. 
The effective area of the flux loop varies during the ramp 
because the fraction of 
the fringe fields overlapping neighbouring dipoles varies.
The saturation of the dipoles, expressed as the change in
effective length, was measured before the LEP
startup on a test stand for different magnet cycles.
The correction between 45 and 90~GeV is of the order
of $10^{-4}$, corresponding to a 5~MeV uncertainty in the physics
energy.

The weak (``10\%'') dipoles matching the LEP arcs to the
straight sections contribute 0.2\% to the total bending field.
Assuming that their field is proportional to that of the main 
bends between RD and physics energies to better than 1\%, their
contribution to any non-linearity in the extrapolation is around 1~MeV.

The bending field of the double strength dipoles in the injection
region contributes 1.4\% of the total. Their bending field has
been measured by additional NMR probes installed in the tunnel in
1998, and found to be proportional to the bending field of the
main dipoles to rather better than $10^{-3}$, which gives a
negligible additional systematic uncertainty.

\section{Quadrupole and horizontal orbit corrector effects}
\label{sec:quad}

\newcommand {\fc}     {f^{\mathrm{RF}}_{\mathrm{c}}}
\newcommand {\fRF}    {f^{\mathrm{RF}}}
\newcommand {\de}     {\Delta E}
\newcommand {\xo}     {X_{0}}
\newcommand {\lh}     {\Delta L_{1}}

\subsection{Earth tides}

The model of earth tides is well understood from LEP1~\cite{ref:zpaper}.
It should be
noted that the amplitude of the tide effect is proportional to energy,
and so is larger at LEP2.

\subsection{Central Frequency and Machine Circumference}
\label{sec:centralf}

For a circular accelerator like LEP the orbit passing on average 
through the centre of the quadrupoles is referred to as the {\it
central orbit}, and the corresponding RF frequency setting is known
as the {\it central RF frequency} $\fc$.  When the RF frequency $\fRF$
does not coincide with $\fc$ the beam senses on average a dipole
field in the quadrupoles, which causes a relative beam energy change $\de$ of :
\begin{equation}
\frac{\de}{E} = - \frac{1}{\alpha} \; \frac{\fRF-\fc}{\fRF}
\end{equation}
where $\alpha$ is the momentum compaction factor, which depends
on the optics used in LEP. Its value is
$1.54 \cdot 10^{-4}$ for the 102/90 optics,
$1.86 \cdot 10^{-4}$ for the 90/60 optics and
$3.86 \cdot 10^{-4}$ for the 60/60 optics, 
with a relative uncertainty of $\lsim 1\%$.

\begin{figure}[t]
\begin{center}
\mbox{\epsfxsize=0.8\textwidth
\epsfbox{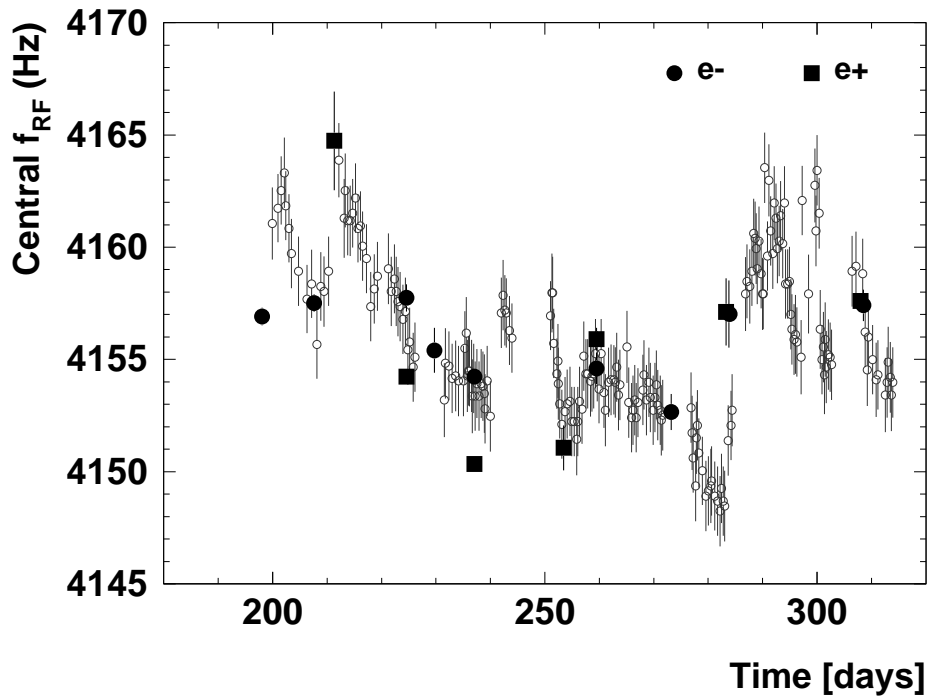} }
\end{center}
\vspace{-0.4cm}
\caption{Evolution of the central frequency as 
a function of time for the 1997 LEP run. The solid points
are actual $\fc$ measurements for $\mathrm{e}^-$ and $\mathrm{e}^+$ 
while the open points 
are obtained from $\xarc$, after correction for
tides. On some occasions,
the electrons and positron $\fc$ differ by up to 4~Hz.
Note that the vertical scale shows a variation in the last four digits of the
LEP RF frequency, which is nominally 352 254 170~Hz.}
\label{fig:fctime}
\end{figure}

The central frequency is only measured on a few
occasions during a year's running and requires non-colliding 
beams~\cite{ref:RFC}. 
The monitoring of the central orbit and of the ring
circumference relies
on the measurement of the average horizontal beam position
in the LEP arcs, $\xarc$~\cite{ref:XARC}. As the length of the beam orbit is
constrained by the RF frequency, a change in machine circumference
will be observed as a shift of the beam position relative to the
beam position monitors (BPMs).
Figure~\ref{fig:fctime} shows the evolution of the 
central frequency determined through $\xarc$ as well as
the direct $\fc$ measurements. The $\xarc$ points have been 
normalised to the electrons $\fc$ measurements. The occasional difference
of about 4~Hz between the electrons and positron $\fc$ 
(measured at physics energy) is not understood.
A similar systematic effect has also been seen in 1998.  
Therefore a systematic error of $\pm$~4~Hz is assigned 
to the central frequency. 
This results in 
an uncertainty of 1.5~MeV in the predicted 
difference between 
the energies measured by RD with the 90/60 and 60/60 optics 
at 45~GeV beam energy.

A correction to the energy of each fill using the measured orbit offset in
the LEP arcs is applied to track the change in $\fc$.

\subsection{RF frequency shifts}

For the first time in 1997, the RF frequency was routinely increased
from the nominal value to change the horizontal damping partition 
number~\cite{ref:design}.
This is a useful technique at high energy, 
causing the beam to be squeezed more in the horizontal plane, and increasing
the specific luminosity, whereas
at lower energy, beam-beam effects prevent
the horizontal beam size reduction. 
Less desirable side effects are that
the central value of the beam energy decreases, the beam 
energy spread increases, and slightly  more RF accelerating voltage is needed
to keep the beam circulating. A 
typical frequency shift of $+100$ Hz gives a beam energy
decrease of about 150~MeV.
Occasionally, when an RF unit trips off, the LEP operators temporarily
decrease the RF frequency to keep the beam lifetime high, in which case
the beam energy values are immediately recalculated instead of 
waiting the usual 15 minutes.

\subsection{Horizontal Corrector Effects}

\label{sec:hcorr}
Small, independently-powered dipole magnets are used to correct
deviations in the beam orbit. 
Horizontal correctors influence the beam energy
either through a change of the integrated
dipole field or through a change of the orbit length $\lh$~\cite{ref:HCORR}.
In general the two effects could be mixed and cannot be 
easily disentangled, although simulations show that the 
orbit lengthening effect
should dominate.
For a given orbit and corrector settings,
the predicted energy shifts can differ by 30\% between the two
models, which implies that a 30\% error should be applied to the energy shifts 
predicted for the corrector settings.

In general the settings of the horizontal correctors
are different for different machine optics, and
also for different beam energies. 
The energy model described by
equation~\ref{eq:model} includes this optics dependent correction
explicitly.
RD measurements using any optics can therefore be combined when
calibrating the NMR probes (section~\ref{sec:nmrpol}), 
which estimate the dominant contribution from the main bend dipoles.

\begin{figure}[t]
\begin{center}
\mbox{\epsfxsize=0.8\textwidth
\epsfbox{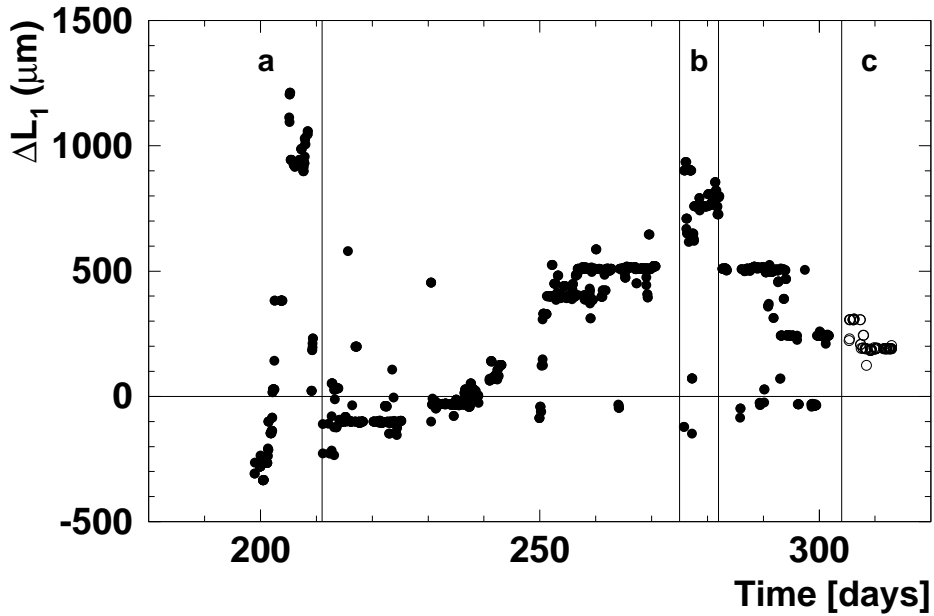} }
\end{center}
\vspace{-0.4cm}
\caption{Evolution of the orbit lengthening
$\lh$ during the 1997 LEP run. In period (a) the beam energy
is 45~GeV, in period (b) 65 and 68~GeV. In period (c)
the 102/90 optics was tested at 91.5~GeV. All other data correspond
to beam energies between 91 and 92~GeV with the 90/60
optics. For $\lh = 500$~$\mu m$, the energy shift is
about 11~MeV at 91.5~GeV with the 90/60 optics.}
\label{fig:l1time}
\end{figure}

For an orbit lengthening $\lh$ the energy change is :
\begin{equation}
\frac{\de}{E} = - \frac{\lh}{\alpha C}
\end{equation}
where $C$ is the LEP circumference and
$\lh$ is calculated from
\begin{equation}
\lh = \sum D_x \delta.
\end{equation}
The sum is over all correctors, $D_x$ is the horizontal
dispersion at the corrector, and $\delta$ is the ``kick'', 
i.e. the deflection due to the corrector. The calculated
and measured values of $D_x$ agree to within 2\%, and
the kick is known from the current in the corrector magnet.
Figure~\ref{fig:l1time}
shows the 
evolution of $\lh$ in physics for the 1997 LEP run.
The size of the effect is somewhat larger than in previous years: 
for a large fraction
of the run $\de$ reaches approximately 11~MeV. 
The contributions of the horizontal correctors to the beam energies
measured by RD for the 90/60 and 60/60 optics differ by about 4~MeV at
45~GeV beam energy.

Recent simulations predict that the  
orbit corrector settings should {\it not} influence
the central frequency by more than 0.5~Hz.
This was confirmed by measurements made during the
1998 LEP run. Separate corrections for
\xarc\ and for the energy shifts due to the
horizontal corrector configurations are therefore applied.

The average beam energy shift from the orbit corrector settings
is 6~MeV for the high energy running in 1997, which is 
much larger than in previous years.
The 30\% model
uncertainty would imply an error of 2~MeV. This is 
increased to
half of the total correction, 3~MeV, in view of the 
evolving understanding of the interplay of 
central frequency and orbit corrector effects.

\subsection{Optics dependent effects}

The majority of the RD measurements were made with the
dedicated polarisation (60/60) optics, while the
physics running was with the 90/60 optics. 
Both horizontal orbit corrector settings (see section~\ref{sec:hcorr}) and 
current differences in the vertically and horizontally 
focussing quadrupoles can cause a difference of a 
few MeV between the beam energies with the two optics. 
These are accounted for by the 
corrections $ C_{\mathrm{h.corr.}}$ and 
$C_{\mathrm{QFQD}} (t)$ in the model of the beam
energy given by equation~\ref{eq:model}. 

Simulations show that the overall 
energy difference can be of either sign, depending on 
the exact imperfections in the machine, and the 
difference is predicted to scale with the beam energy. The predicted
difference at 45~GeV also has an uncertainty of 1.5~MeV from 
central frequency effects, described in section~\ref{sec:centralf}. 
The measured difference in the data is evaluated from 
the residuals $\Epol - \Enmr$ with respect to the
simultaneous fit to all RD measurements, using either optics.
These can be seen in figure~\ref{fig:allresid}(a).
The observed average beam energy difference between
RD measurements with the two optics is:
\beq E(90/60) - E(60/60) = +2 \mbox{ MeV at 45 GeV.}\eeq 
From the fill-to-fill scatter, the uncertainty on this measured difference
is $ \le 1$~MeV.
Scaling the difference with the beam energy, 
a systematic uncertainty of 4~MeV is therefore taken to
cover all uncertainties due to 
optics dependent effects at physics energy.

\section{Evaluation of the centre-of-mass energy at each IP}
\label{sec:ipcor}

As at LEP1, corrections to the centre-of-mass energy arise from the non-uniformity
of the RF power distribution around LEP and from possible offsets of the beam
centroids during collisions in the presence of opposite-sign vertical dispersion\cite{ref:zpaper}.

\subsection{Corrections from the RF System}
\begin{figure}[htb]
\centering
\mbox{\epsfxsize=0.6\textwidth
\epsfbox{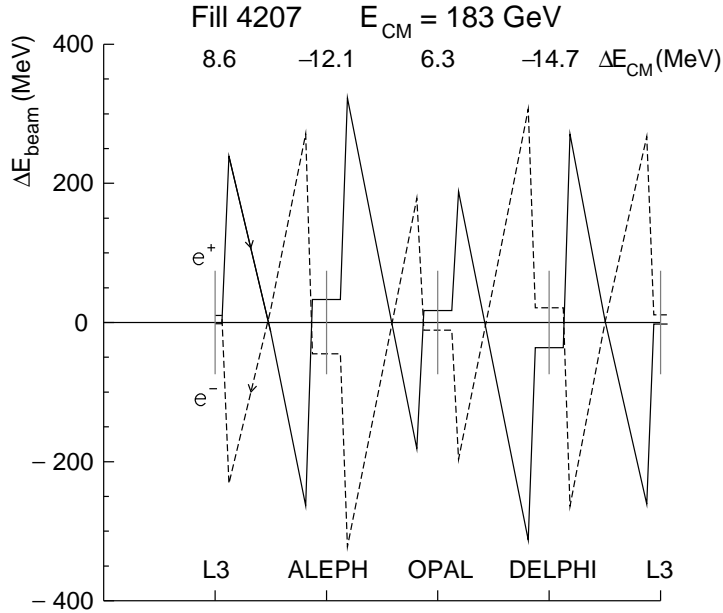} }
\caption{The evolution of the energy of each of the counter-rotating
beams as they circulate in LEP, the so-called {\it energy sawtooth}.  
The electron beam is represented by the left-going
dotted line,  the positron beam by the right-going solid line.  The light grey lines
mark the positions of the IPs.
$\Delta E = 0$ denotes the average energy
of LEP,   i.e.  that given by the rest of the energy model.  
The energy gains provided by the RF are
clearly visible.  
This represents a typical RF configuration and
centre-of-mass energy corrections for the 1997 running.
$\Delta \ECM$ is the shift in \ECM\ at each IP due to RF effects.}
\label{fig:sawtooth}
\end{figure}

Since the beam energy loss due to synchrotron radiation is
proportional to $E_{\mathrm {beam} }^4$, operation of LEP2 requires a
large amount of RF acceleration to maintain stable beam orbits.  To
provide this acceleration, new super-conducting (SC) RF cavities have
been installed around all of the experiments in LEP.   This implies
that, contrary to LEP1, the exact anti-correlation of RF effects
on the beam energy at IP4 and IP8 is no longer guaranteed, and that
large local shifts in the  beam energy can occur at any of the IPs. 
This also implies that the energy variation in the beams (the ``sawtooth'') as they circulate
around LEP is quite large (see figure \ref{fig:sawtooth}.).
This increases the sensitivity of the centre-of-mass
energy to non-uniformities in the energy loss arising from differences in the local
magnetic bend field, machine imperfections, etc.

The modelling of the energy corrections from the RF system
is carried out by the iterative calculation of the stable RF phase angle $\phi_s$
which proceeds by setting the total energy gain,
$V_{\mathrm{tot}} \sin \phi_s$, of the 
beams as they travel around the machine equal to all of the known energy losses.
Here $V_{\mathrm{tot}}$ is the total RF accelerating voltage.
The measured value of the synchrotron tune $Q_s$ and the energy offsets between the
beams as they enter and leave the experimental IPs are used to constrain energy
variations due to overall RF voltage scale and RF phase errors.  In particular, the
phase error at each IP is set using the size of the energy sawtooth measured by the LEP BOM system
compared with the total RF voltage at that IP.  This is a powerful constraint on potential
phasing errors, as was seen in the LEP1 analysis.

The average corrections for the 1996 and 1997 running are shown in 
table~\ref{corrs}.  The corrections for running below the WW threshold are typically smaller
by a factor of two, with correspondingly smaller errors.
The bunchlet-to-bunchlet variation in corrections is negligible.  

\begin{table}[tbh]
\begin{center} 
\begin{tabular}{c|rrr} \hline
&    \multicolumn{3}{c}{$\Delta \ECM$ (MeV)} \rule[0mm]{0mm}{5mm}\\
LEP IP &\hfil $\ECM =161$ GeV\hfil &\hfil $\ECM =172$ GeV\hfil &\hfil $\ECM =183$ GeV\hfil  \rule[0mm]{0mm}{5mm} \\  \hline

2      & 19.7\hspace*{.8cm} & 18.8\hspace*{.8cm} & 8.1\hspace*{.8cm}\rule[0mm]{0mm}{5mm}\\ 
                         
4      & --5.5\hspace*{.8cm} & --5.8\hspace*{.8cm} & --10.8\hspace*{.8cm}\rule[0mm]{0mm}{5mm}\\ 
                         
6      & 20.2\hspace*{.8cm} & 19.5\hspace*{.8cm} & 5.9\hspace*{.8cm}\rule[0mm]{0mm}{5mm}\\ 
                         
8      & --9.3\hspace*{.8cm} & --8.0\hspace*{.8cm} & --13.3\hspace*{.8cm}\rule[0mm]{0mm}{5mm}\rule[-5mm]{0mm}{5mm}\\
\hline
\end{tabular}
\caption{ The average corrections (in MeV) to $\ECM$ for each LEP IP and each 
energy point for the 1996 and 1997 RF models.\label{corrs}}
\end{center}
\end{table}

As at LEP1, the errors on the energy corrections are evaluated by a comparison
of those quantities ($Q_s$, the orbit sawtooth, and the longitudinal position of the
interaction point)
which can be calculated in the RF model and
can be measured in LEP.  In addition, uncertainties from the inputs to
the model, such as the misalignments of the RF cavities and the
effects of imperfections in the LEP lattice, must also be considered.
Since many of the uncertainties on the RF corrections scale with
the energy loss in LEP, however, the overall uncertainty due to RF effects
is larger at LEP2 than at LEP1.  Note that the errors are given below
in terms of $\Ebeam$, and are obtained by dividing the error on $\ECM$ by two.

Comparison of the measured and calculated $Q_s$ values reveals a
discrepancy in the modelled and measured overall RF voltage seen by the beam.
The difference is small, on the order of 4\%, which can be explained by
an overall scale error in the measured voltages or a net phase error in
the RF system of a few degrees.  An overall scale error changes the energy
corrections by a corresponding amount ({\it i.e.}, a 10~MeV correction acquires
an error of 0.4~MeV, which is negligible), whereas phase errors can shift the
energy by larger amounts at the IP closest to the error.
For 1996, the overall error due to this
mismatch was computed by assuming the entire phase error was localised to one
side of an IP, and the largest shift taken as the error for all IPs (4~MeV $\Ebeam$).  
For 1997,
the total energy gain was normalised so that $Q_s$ was correct, and the phase
errors for each IP were calculated using the orbit measurement of the local
energy gain.  This resulted in a smaller error of 1.5~MeV on $\Ebeam$ from the voltage scale
and phasing effects. 

The positions of all of the RF cavities in LEP have been measured repeatedly
using a beam-based alignment technique with a systematic precision of 1 mm and
a 1 mm rms scatter over time~\cite{ref:RFalign}.
The systematic error on the energy corrections
is evaluated by coherently moving the RF cavities in the model 
by 2 mm away from (towards)
the IPs, and observing the change in $\ECM$.  This results in a 1.5~MeV error
on $\Ebeam$ for 1996 and 1997.

Recently, a study of the effects of imperfections in the LEP lattice on 
the energy loss of the beams at LEP2 has been performed \cite{ref:jjowettRF}.
Calculations of the centre-of-mass energy in an ensemble of machines with
imperfections similar to those of LEP yields an rms spread of 2.5~MeV $\Ebeam$
in the predicted energy at the IPs due to non-uniformities in the energy loss
of the beams.  These shifts only depend on the 
misalignments and non-uniformities of all of the magnetic
elements in LEP, which are essentially
unmeasurable, and is not contained in any of the
other error sources.

In order to keep the error estimate as conservative as possible,
the error on the energy corrections from the RF should be considered 
100\% correlated between IPs and energy points.  The total error from
RF effects for each energy point is given in table~\ref{tab:error}.

\subsection{Opposite sign vertical dispersion}

In the bunch train configuration, beam offsets at the collision point
can cause a shift in the centre-of-mass energy due to opposite sign
dispersion~\cite{ref:zpaper}. 
The change in energy is evaluated from the calculated
dispersion and the measured beam offsets from beam-beam deflection scans.

The dispersions have been calculated 
using MAD~\cite{ref:mad} for all the configurations in 1996 and 1997. 
No dedicated measurements of dispersion were made in 1996, while 
in 1997 the measured values agree with the prediction to 
within about 50\%. This is the largest cause of
uncertainty in the possible correction.
Beam offsets were
controlled to within a few microns by beam-beam deflection scans. 
The resulting luminosity-weighted correction to the centre-of-mass 
energies are typically 1 to 2~MeV, with an error of about 2~MeV.
No corrections have been applied 
for this effect, and an uncertainty of 2~MeV has been assigned.

\section{Summary of systematic uncertainties}
\label{sec:syst}

\begin{table}[htb]
\begin{center}
\begin{tabular}{l|c}
\hline
Source                                     & Error [MeV]  \\
\hline
 Extrapolation from NMR--polarisation:      &    \\
\hspace*{0.4cm} NMR rms/$\sqrt{N}$ at physics energy          & 10 \\
\hspace*{0.4cm} Different \Epol\ fills                     & 5  \\
\hline
 Flux-loop test of extrapolation:           &               \\
\hspace*{0.4cm} NMR flux-loop difference at physics energy& 20 \\
\hspace*{0.4cm}  Field not measured by flux loop           & 5  \\
\hline
Polarisation systematic                    & 1 \\
$\ee$ energy difference                    & 2 \\
Optics difference                          & 4 \\
\hline
Corrector effects                          & 3  \\
Tide                                       & 1  \\
Initial dipole energy                      & 2  \\
Dipole rise modelling                      & 1  \\
\hline
IP specific corrections ($\delta \ECM/2$):&    \\
\hspace*{0.4cm} RF model                                   & 4  \\
\hspace*{0.4cm} Dispersion                                 & 2  \\
\hline
Total                                      & 25 \\
\hline
\end{tabular}
\end{center}
\caption{Summary of contributions to the 1997 beam energy 
uncertainty.}
\label{tab:error}
\end{table}

The contributions from each source of uncertainty
described above are summarised in table~\ref{tab:error}.
The first groups describe the uncertainty in the normalisation
derived from NMR-polarisation comparisons, NMR-flux-loop tests and
the part of the bending field not measured by the flux-loop. These
extrapolation uncertainties dominate the analysis.
The subsequent errors concern the 
polarisation measurement, specifically its intrinsic 
precision (which is less than 1~MeV),
the possible difference in energy between electrons
and positrons, and the difference between optics.
None of the additional uncertainties  from 
time variations in a fill, and IP specific  
corrections contribute an uncertainty greater than 5~MeV. 

\subsection{Uncertainty for data taken in 1996}
\label{sec:e96err}

The analysis of the 1996 data was 
largely based on a single fill with RD measurements 
at 45 and 50~GeV. The apparent consistency
of the flux-loop and NMR data compared to RD data 
was about 2~MeV over this 5~GeV interval, 
i.e.\ a relative error $4 \times 10^{-4}$, which using
a naive linear extrapolation 
would give an uncertainty of 13.5 (15)~MeV at 81.5 (86)~GeV. 
These errors were inflated to 27 (30)~MeV before the 1997 data were 
available, since there was no test of reproducibility from fill to fill, 
there was no check of the non-linearity possible from a fill
with two energy points, and the field outside the flux loop had
not been studied.
Although more information is available in 1997, this 
larger uncertainty is retained
for the 1996 data, partly motivated by 
the sparsity of RD measurements in that year. In addition,
the single parameter fit that was used for 1996
leads to a shift of 20~MeV, and an increased scatter of 60~MeV,
when used to predict the energies in physics in 1997.

The uncertainties for the two 1996 data samples can be assumed to
be fully correlated. However, the extrapolation uncertainty for the
1997 data is somewhat better known. Since the energy difference 
between the maximum RD energy and the physics energy is
nearly the same in the two years, it can be assumed that
the 25~MeV uncertainty of 1997 data is common to the 1996 data.

\subsection{Lower energy data taken in 1996 and 1997}

During 1996 and 1997, LEP also operated at the Z resonance,
to provide data samples for calibrating the four experiments,
and at intermediate centre-of-mass energies, 130--136~GeV,
to investigate effects seen at the end of 1995 at ``LEP 1.5''.
The dominant errors on the beam energy 
are from the extrapolation uncertainty, and
scale with the difference between physics energy and RD energy.
The optics difference scales in the same way. 
Several other effects such as the tide correction
are proportional to the beam energy. The dipole rise
per fill depends in addition on whether bend modulation was 
carried out at the start of fill. The total beam energy uncertainties 
are found to be 6~MeV for Z running, and 14~MeV for
LEP 1.5 running.

\section{Centre-of-mass energy spread}
\label{sec:espread}

\begin{table}[t]
\begin{center}
\begin{tabular}{ccr|rr }
\hline
Year & $\ECM$ [GeV] & $\cal{L}$ [\ipb] &
\multicolumn{2}{c}{ $\sigma _ {\Ebeam}$ [MeV] } \\
 & &  & Predicted & Derived  \\
\hline
1996 & 161.3 &  10      &  $102 \pm 5$&  $105\pm  1 \pm 5  $\\
     & 164.5 & 0.05     &  $106 \pm 5$&  $ 97\pm 35 \pm 4  $\\
\hline 
     &  All ``161''& 10 &  $102 \pm 5$&  $105  \pm5$    \\
\hline
     & 170.3 & 1        &  $115 \pm 6$ &  $125 \pm 4 \pm 4  $\\
     & 172.3 & 9        &  $117 \pm 6$ &  $111 \pm 1 \pm 4  $\\
\hline
     & All ``172'' & 10 &  $117 \pm 6$ &  $113 \pm 4 $    \\
\hline
1997 & 130.0 & 3        &  $ 66 \pm 3$ &  $ 71 \pm 1 \pm 4  $\\
     & 136.0 & 3        &  $ 74 \pm 4$ &  $ 80 \pm 1 \pm 4  $\\
\hline
     & 180.8 & 0.2      &  $141 \pm 7$ &  $106 \pm 14 \pm 3 $\\
     & 182.0 & 6        &  $164 \pm 8$ &  $146 \pm 5 \pm 5  $\\
     & 182.7 & 46       &  $154 \pm 8$ &  $154 \pm 6 \pm 6  $\\
     & 183.8 & 2        &  $145 \pm 7$ &  $124 \pm 4 \pm 6  $\\
\hline
     & All ``183'' & 54 &  $155 \pm 8$ &  $152\pm8$    \\
\hline
\end{tabular}
\end{center}
\caption{Beam energy spreads. Note that these should be 
multiplied by $\sqrt{2}$ to give the centre-of-mass
energy spread. 
The predicted values, and the values derived from the 
bunch length measurement are given, together with the 
approximate luminosity, $\cal{L}$, recorded at each nominal centre-of-mass
energy.
Luminosity weighted
predictions for all fills with centre-of-mass energy 
close to 161, 172 and 183~GeV are also given.}
\label{tab:espread}
\end{table}

The spread in centre-of-mass energy is relevant for evaluating the 
width of the W boson, which is about 2~GeV. 
The beam energy spread can be predicted for a particular optics,
beam energy and RF frequency shift. This spread has been 
calculated for every 15 minutes of data taking, or more
often in the case of an RF frequency shift. 
Weighting the prediction by the (DELPHI) integrated luminosity gives the
average ``predicted'' values in table~\ref{tab:espread} for 
each nominal centre-of-mass energy. Overall averages for
all data taken close to 161, 172 and 183~GeV are also listed. The
error in the prediction is estimated to be about 5\%, from
the differences observed when a quantum treatment of
radiation losses is implemented.

The beam energy spread can also be derived from the 
longitudinal bunch size measured by one of the experiments. This 
procedure has been applied to the
longitudinal size of the interaction region measured
in ALEPH, $\sigma_z^{\mathrm{ALEPH}}$, 
which is related to the energy spread by~\cite{ref:zpaper,ref:espread}:
\beq
\sigma _ {\Ebeam} = \frac {\sqrt{2} \Ebeam} 
{\alpha R_{\mathrm{LEP}} } \Qs \sigma_z^{\mathrm{ALEPH}}.
\eeq
The momentum compaction factor $\alpha$ is
known for each optics, $R_{\mathrm{LEP}}$ is 
the average radius of the LEP accelerator, and
$\Qs$ is the incoherent synchrotron tune. 
This is derived from the measured coherent \Qs\ using:
\beq
\frac {\Qs^{\mathrm{coh}}}{\Qs^{\mathrm{incoh}}} = 1 - \kappa
\frac{I^{\mathrm{bunch}}} { 300 \mu \mathrm{A}}
\eeq
The parameter $\kappa$ was measured in 1995 to be $0.045\pm0.022$
at the Z. For the same $\Qs$ and machine configuration, this
would scale with $1/\Ebeam$. This scaling has been used for the
central values of $\sigma _ {\Ebeam}$ evaluated from the measured
bunch lengths in table~\ref{tab:espread}. However, the reduction
in the number of copper RF accelerating cavities in the machine
since 1995
is expected to further reduce the value of $\kappa$, so the
uncertainty of $\pm 0.022$ is retained for all energies. The value
and uncertainty are consistent with estimates from 
the variation of bunch length with current measured with the
streak camera\footnote{The streak camera 
measures the bunch length parasitically
by looking at synchrotron light emitted when the bunch
goes through a quadrupole or wiggler magnet.}
in 1998. 
Where two errors are quoted for
the derived number, they are the statistical
uncertainty in the bunch length measurement, and a
systematic uncertainty, which is dominated by the
uncertainty in $\kappa$, with a 1~MeV contribution from
the uncertainty in $\alpha$. The predicted and derived values
agree well within the quoted errors.

The measurement of \Qs\ is difficult for high energy beams, 
and in 1997, a reliable value is only available for 58\% of the 
data. 
It is therefore recommended to use the predicted values. 
The beam energy spreads must be multiplied by 
$\sqrt{2}$ to
give the centre-of-mass energy spreads~\cite{ref:zpaper,ref:espread}, which are:
$144\pm7$~MeV at 161~GeV, $165\pm8$~MeV at 172~GeV and
$219\pm11$~MeV at 183~GeV.

\section{Conclusions and outlook}
\label{sec:conclude}

The method of energy calibration by magnetic extrapolation of
resonant depolarisation measurements at lower energy has made
substantial progress with the 1997 data. The success
in establishing polarisation above the Z has allowed a 
robust application of the method, and 
the mutual consistency of the resonant depolarisation, 
NMR and flux-loop  data
has been established at the 
20~MeV level at physics energy, with a total systematic
uncertainty in the beam energy of 25~MeV. 
The precision is limited by the understanding of the
NMR/flux-loop comparison.

As LEP accumulates more high energy data, the experiments
themselves will be able to provide a cross-check on the centre-of-mass
energy by effectively measuring the energy of the emitted photon in
events of the type $e^+ e^- \to {\rm Z} \gamma \to f \bar f \gamma$,
where the Z is on-shell.  This can be done using a kinematic fit
of the outgoing fermion directions and the precisely determined Z-mass
from LEP1.  The ALEPH collaboration have shown\cite{ref:aleph_ebeam}
the first attempt to make this measurement in the $q\bar q \gamma$ channel,
where they achieve a precision of 
$\delta\Ebeam = \pm 110 (\mathrm{stat}) \pm 53 (\mathrm{syst})$~MeV.  With 500 pb$^{-1}$
per experiment, the statistical precision on this channel should approach 15~MeV.
Careful evaluation of systematic errors will determine the usefulness of this approach.

In future, a new apparatus will be available for measuring
the beam energy.
The LEP Spectrometer Project~\cite{ref:massimo}
will measure the bend angle of the beams using standard LEP
beam pick ups with new electronics to measure the position
to the order of a micron precision as they enter or leave
a special dipole in the LEP lattice whose bending field
has been surveyed with high precision. A first phase
of the spectrometer is already in place for the 1998 running,
with the aim of checking the mechanical and thermal stability
of the position measurement. In 1999, the new magnet will
be installed, and the aim is to use this new, independent
method to measure the beam energy to 10~MeV at high energy.
It should be possible to propagate any improvement in
the beam energy determination back to previous years
by correcting the extrapolation and correspondingly
reducing the uncertainty.

\section*{Acknowledgements}

The unprecedented performance of the LEP collider in this new
high energy regime is thanks to the SL Division of CERN. 
In particular, careful work and help of many people in SL Division
has been essential in making specific measurements
for the energy calibration.

We also acknowledge the support of the  
Particle Physics and Astronomy Research Council,  UK.


\begin{thebibliography}{9}
\bibitem{ref:zpaper}
``Calibration of centre-of-mass energies at LEP1 for precise
measurements of Z properties'', 
LEP Energy Working Group,  Eur.\ Phys.\ J.\ C 6 (1999) 2, 187-223.
\bibitem{ref:thomson}
See for example papers by
H. Przysiezniak and M. Thomson,
ICHEP98, Vancouver, Canada, 22-29 July 1998.
\bibitem{ref:kmod}
B. Dehning et al.,
``Dynamic beam based calibration of beam position monitors'',
CERN-SL-98-038-BI, June 1998.
        Presented at 6th European Particle Accelerator Conference (EPAC 98),
        Stockholm, Sweden, 22-26 June 1998. 
\bibitem{ref:jan}
J.\ A.\ Uythoven, 
``A LEP (60,60) Optics for Energy Calibration Measurements'',
CERN-SL-97-058-OP.
\bibitem{ref:bernd}
See for example
B. Dehning in the proceedings of the 8th LEP 
Performance Workshop, Chamonix, 1998,
CERN-SL-98-006 DI.
\bibitem{ref:RFC} R. Bailey et al., ``LEP Energy Calibration'',
Proc. of the 2nd EPAC, Nice, France (1990) and CERN SL/90-95. \\
 H. Schmickler, ``Measurement of the Central Frequency
of LEP'', CERN SL-MD note 89 (1993).
\bibitem{ref:XARC} J. Wenninger, ``Radial Deformations of the LEP ring'',
CERN SL / Note 95-21 (OP). \\
J. Wenninger, ``Measurement of Tidal Deformations of the LEP
Ring with Closed Orbits'', SL-Note 96-22 (OP).
\bibitem{ref:HCORR} J. Wenninger, ``Orbit Corrector Magnets and Beam Energy'',
SL-Note 97-06 OP.
\bibitem{ref:design}
LEP Design Report Vol. III, LEP2, CERN-AC/96-01(LEP2).
\bibitem{ref:RFalign}
LEP Energy Working Group Note
97-03, ``A Systematic Check of SC Cavity Alignment Using LEP Beams'', 
M.D. Hildreth.
\bibitem{ref:jjowettRF}
J.M. Jowett, ``Monte-Carlo Study of the (102$^\circ$, 90$^\circ$) Physics Optics for LEP'', 
SL-Note-97-84 AP.
\bibitem{ref:mad}
H. Grote and F.C. Iselin,  ``The MAD program'',  version 8.16,  
CERN SL/90--13  (AP)  (rev.4,  March 19,  1995) 
\bibitem{ref:espread}
LEP Energy Working Group Note 
96-07, ``Determination of the LEP Energy Spread Using Experimental Constraints'', 
E. Lan\c{c}on and A. Blondel.
\bibitem{ref:aleph_ebeam} 
The ALEPH Collaboration, ``Preliminary Evaluation of the LEP Centre-of-Mass
Energy Using Z$\gamma$ Events'', Contributed paper 1038 to
ICHEP98, Vancouver, Canada, 22-29 July 1998.
\bibitem{ref:massimo}
See for example
M. Placidi in the proceedings of the 8th LEP 
Performance Workshop, Chamonix, 1998,
CERN-SL-98-006 DI.
\end{thebibliography}
\end{document}